\documentclass[twocolumn,secnumarabic,amssymb, nobibnotes, aps, prd]{revtex4-2}
\usepackage{natbib}
\usepackage{graphics}
\usepackage{multirow}

\newcommand{\env}[1]{\texttt{#1}}
\begin{document}
\raggedbottom
\setlength{\parskip}{0pt}
\graphicspath{ {./} }

\begin{abstract}
Reflection high-energy electron diffraction (RHEED) is a ubiquitous in situ molecular beam epitaxial (MBE) characterization tool. Although RHEED can be a powerful means for crystal surface structure determination, it is often used as a static qualitative surface characterization method at discrete intervals during a growth. A full analysis of RHEED data collected during the entirety of MBE growths is made possible using principle component analysis (PCA) and k-means clustering to examine significant boundaries that occur in the temporal clusters grouped from RHEED data and identify statistically significant patterns. This process is applied to data from homoepitaxial SrTiO$_{3}$ growths, heteroepitaxial SrTiO$_{3}$ grown on scandate substrates, BaSnO$_{3}$ films grown on SrTiO$_{3}$ substrates, and LaNiO$_{3}$ films grown on LaAlO$_{3}$ substrates. This analysis may provide additional insights into the surface evolution and transitions in growth modes at precise times and depths during growth, and that video archival of an entire RHEED image sequence may be able to provide more insight and control over growth processes and film quality.
\end{abstract}
\title{Machine Learning Analysis of Perovskite Oxides Grown by Molecular Beam Epitaxy}%
	
\author{Sydney R. Provence}
\email{srp0046@auburn.edu}
\affiliation{Department of Physics, Auburn University, Auburn, Alabama 36830, USA}
\author{Suresh Thapa}
\affiliation{Department of Physics, Auburn University, Auburn, Alabama 36830, USA}
\author{Rajendra Paudel}
\affiliation{Department of Physics, Auburn University, Auburn, Alabama 36830, USA}
\author{Ryan B. Comes}
\email{ryan.comes@auburn.edu}
\affiliation{Department of Physics, Auburn University, Auburn, Alabama 36830, USA}
\author{Tristan Truttmann}
\affiliation{Department of Chemical Engineering and Materials Science, University of Minnesota, Minneapolis, Minnesota 55455, USA}
\author{Abhinav Prakash}
\affiliation{Department of Chemical Engineering and Materials Science, University of Minnesota, Minneapolis, Minnesota 55455, USA}
\author{Bharat Jalan}
\affiliation{Department of Chemical Engineering and Materials Science, University of Minnesota, Minneapolis, Minnesota 55455, USA}
\date{March 2020}%
\maketitle
	
\section{Introduction}
Reflection high energy electron diffraction (RHEED) is one of the most ubiquitous tools for in situ analysis of films growth by molecular beam epitaxy (MBE). The basic implementation of RHEED involves an electron gun positioned at grazing incidence to scatter electrons off a single crystal substrate. Electrons are diffracted onto a phosphor screen, creating a pattern of high-intensity streaks and spots from scattering. The shallow penetration depth of the electron gun makes RHEED predominantly sensitive to the first few surface layers of the crystallographic structure \cite{ichimiya2004reflection}. As such, the images derived from RHEED patterns can be considered real-time measurements of the properties of the crystalline surface during epitaxial growth. 

RHEED patterns contain both qualitative and quantitative information about a growth, such as the in-plane lattice parameters \cite{matolin1995subpixel}, growth mode \cite{nikiforov2000situ,song2003srtio3} and surface disorder \cite{larsen1990influence}. The intensity oscillations in RHEED patterns during growth have been commonly used to control the film thickness during epitaxial growth, as the periodicity of the oscillation is correlated with the growth rate \cite{neave1983dynamics, heyn1997correlation}. In typical single or two-component materials such as Si or GaAs, the oscillation period directly corresponds with the deposition time of a single monolayer in a layer-by-layer growth mode, and thus is equivalent to the growth rate measured in monolayers/second. In multicomponent complex oxides such as SrTiO$_{3}$ (STO), it has been demonstrated that changes in the surface reconstruction and RHEED intensity can serve as a measure of stoichiometry, although the use of oscillations as a measure of the completeness of a full layer is less well understood \cite{kajdos2014surface, haeni2000rheed, sun2018chemically}.

Despite the relative wealth of information contained in RHEED patterns and the near universality of the presence of RHEED in both commercial and academic MBE chambers, the analysis of RHEED is typically limited to a quantitative analysis of a few static images taken before, during, and after the growth and/or the mean intensity collected of a few pre-determined specular or diffraction spots over the course of a growth. Although systems that allow users to take video of the evolution of RHEED patterns have been developed and are commercially available \cite{gur1997system, barlett1991ccd}, video is rarely used in RHEED analysis beyond the analysis of a few static frames and the majority of the information contained within goes unused.

Methods for analysis of the full RHEED image sequence using machine learning techniques have been proposed by Vasudevan et al \cite{vasudevan2014big}. Machine learning algorithms can be subdivided into two classes, supervised and unsupervised learning \cite{james2013statistical}. Algorithms in the former method use data where the output values are already known, while the latter method attempts to discern structure from unlabeled data points. Unsupervised learning methods have been demonstrated in materials analysis applications such as scanning probe microscopy \cite{kalinin2016big, belianinov2015big}, scanning transmission electron microscopy \cite{jesse2016big}, transport measurements \cite{strelcov2014deep}, and crystal structure predictions \cite{takahashi2019creating}. Vasudevan et al \cite{vasudevan2014big} have demonstrated an unsupervised learning approach for the interpretation an entire sequence of RHEED data that utilizes principle component analysis (PCA) and a k-means clustering algorithm to identify the areas in the RHEED pattern with the most statistical variance and identify transitions in the growth mode. In this study, we expand upon these initial results to employ machine learning to interpret film stoichiometry, growth modes, strain relaxation, surface termination, and crystallinity in MBE-grown films.

\section{Methods}
We apply the machine learning approach to RHEED videos collected during the MBE growth of homoepitaxial STO films, heteroepitaxial STO films grown on GdScO$_{3}$ (GSO) and TbScO$_{3}$ (TSO) substrates, heteroepitaxial BaSnO$_{3}$ (BSO) films grown on STO substrates, and the shuttered deposition of LaNiO$_{3}$ (LNO) on LaAlO$_{3}$ (LAO) substrates. These film case studies allow us to observe a variety of growth dynamics to correlate thin film evolution with observed machine learning trends in the RHEED data. Details of the film growth conditions and MBE configuration for all samples are provided in the supplementary material. RHEED patterns for STO and LNO films were recorded along the [110] azimuth using a kSA 400 acquisition system, and video of the RHEED patterns during growth was saved using FlashBack screen recorder. Video was acquired along the [100] azimuth for the BSO films and were collected with an EZRHEED by MBE Control Solutions. 

The general workflow for video processing in our approach is based on the process proposed by Vasudavan et al \cite{vasudevan2014big}, involving (i) cropping and decomposing videos into individual greyscale frames, (ii) applying PCA to compress the total size of the individual frames, and (iii) applying a k-means clustering algorithm to the constructed feature space to group frames temporally. The application of PCA compresses the data, so that each frame is represented by a linear combination of \(D\) principle components (eigenvectors) and their associated time-dependent loadings (eigenvalues). Every frame in the sequence can be reconstructed by multiplying the principle components by the loadings. We find that using \(D=5\) allows for the full image sequence to be restored while retaining over 95\% of the variance of the initial dataset. K-means is an iterative clustering algorithm that breaks the frames into groups in which each individual frame is grouped into the cluster with the nearest mean image. Naive K-means analysis requires that the number of clusters, \(K\), is pre-determined by the user before running the algorithm, which can often make determining the most ``natural'' number of clusters difficult. A more detailed procedure for the analysis is given in the supplementary material, including links to the source code for other groups to implement in their work.

\section{Results and Discussion}
\subsection{STO Homoepitaxial Growth}
\begin{figure}
	\includegraphics{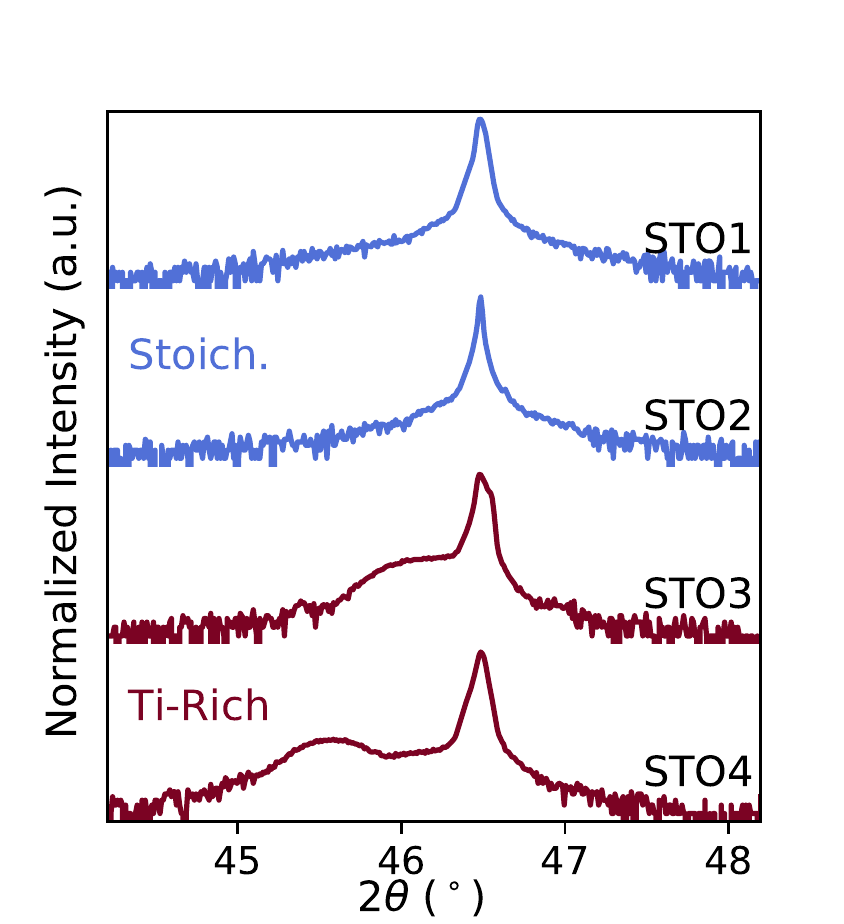}
	\caption{\label{XRD}XRD 2$\theta$ curves of the STO heteroepitaxial samples. The curves of samples STO1 and STO2 indicate that the samples are approximately stoichiometric, while STO3 and STO4 are off stoichiometry.}
\end{figure}

Four STO films were grown on $\langle$100$\rangle$ STO substrates with the substrate thermocouple heated to 1000$^{\circ}$C and a variable ratio of TTIP outlet pressure to Sr flux (Table \ref{STOsamples}). The growth rate for each film is estimated to be 0.02 u.c./s with a total thickness of approximately 26 nm. X-ray diffraction (XRD) scans on the samples indicate that samples STO1 and STO2 were stoichiometric films, while STO3 and STO4 were non-stoichiometric (Fig. \ref{XRD}). X-Ray Photoelectron Spectroscopy (XPS) data for each sample in tandem with the XRD scans indicate that samples STO3 and STO4 are both titanium-rich. For more details on the film growth and XRD and XPS data from these samples, refer to Thapa et al \cite{thapa2020}.

\begin{table}
\caption{\label{STOsamples}STO-STO Samples Ti:Sr XPS Ratios.}
\begin{ruledtabular}
\begin{tabular}{c|c|c}
Sample & Ti2p:Sr3d (XPS Surface) & Ti2p:Sr3d (XPS Normal)\\
\hline
STO1 & 0.6 & 0.628\\
STO2 & 0.617 & 0.647\\
STO3 & 0.775 & 0.872\\
STO4 & 1.144 & 1.165\\
\end{tabular}
\end{ruledtabular}
\end{table}

\begin{figure*}
	\includegraphics{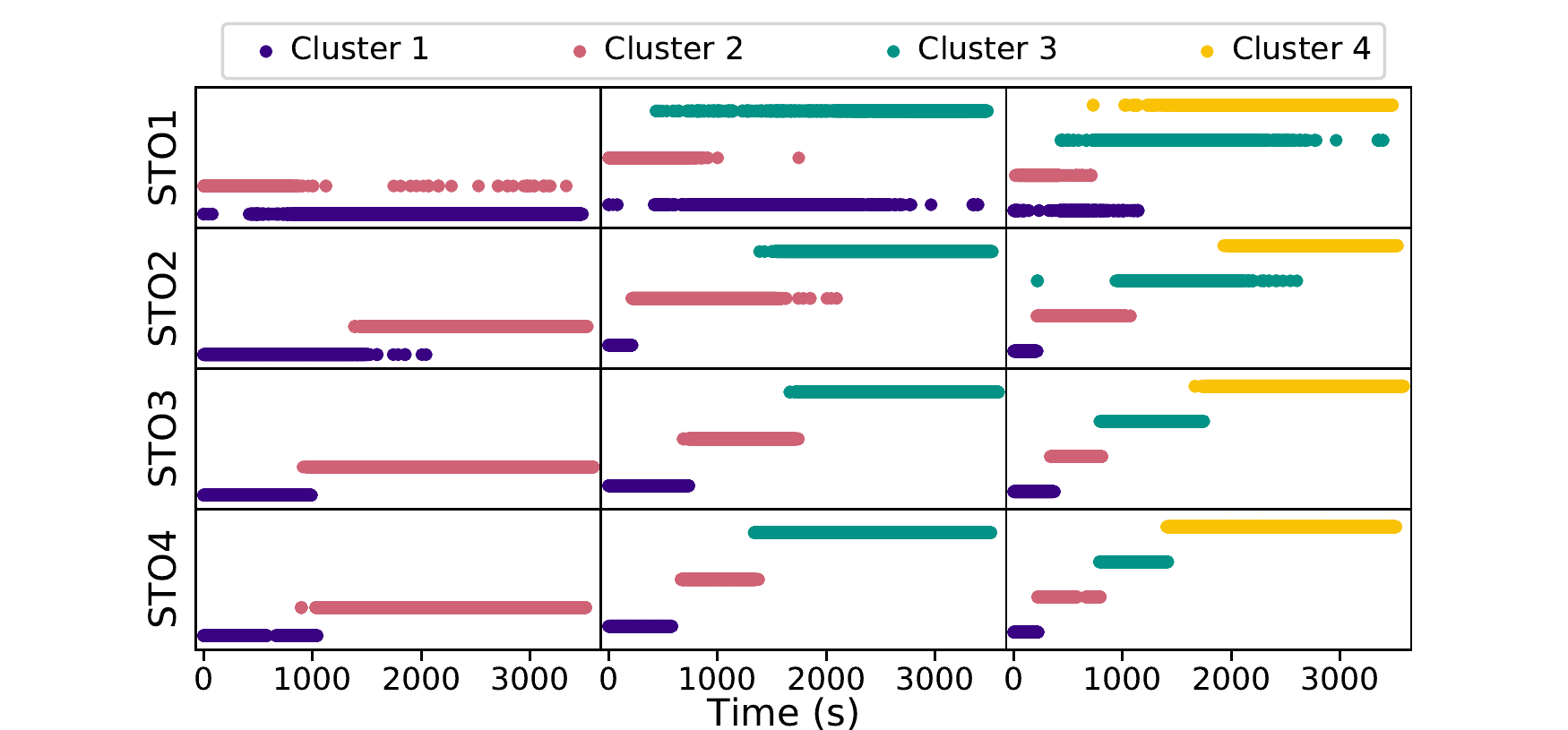}
	\caption{\label{STOClustering}K-means clustering as applied to videos of the deposition of STO on STO substrates with varied TTIP:Sr flux ratios using two (left column), three (center), and four clusters (right). The non-stoichiometric samples, STO3 and STO4, have clear time boundaries in the clusters of frames of the video, suggesting clear stages in the surface evolution during growth. The cluster boundaries are less clear in stoichiometric samples STO1 and STO2.}
\end{figure*}

K-means clustering for each frame in the RHEED video for each sample is plotted in Fig.~\ref{STOClustering} using \(K=2, 3\), and \(4\) clusters. The stoichiometric samples lack obvious boundaries between each cluster, which intuitively makes sense—if there is no obvious change in the RHEED pattern over the course of the growth, as would be expected for a sample in which the film quality and stoichiometry mirrors that of the substrate, the algorithm will attempt to cluster frames based on intensity fluctuations due to vibrations in the equipment or RHEED pattern intensity rather than actual pattern shifts. This effect is most apparent in STO1, in which the clusters tend to share frames temporally (Table~\ref{STOClusterDuration})) and lack distinct boundaries, although a distinct initial cluster appears to form for higher values of K for the first $\sim$1000 seconds (see Thapa et al \cite{thapa2020} for additional details). For STO2, a single distinct cluster develops in the initial 213 seconds of growth when the frames are clustered using \(K > 3\), although the boundaries between other groupings are non-distinct. 

\begin{table*}
	\caption{\label{STOClusterDuration}STO-STO Samples Cluster Durations.}
	\begin{ruledtabular}
		\begin{tabular}{c|cc|ccc|cccc}
			\multirow{2}{*}{Sample} & \multicolumn{2}{c|}{\(K=2\)} & \multicolumn{3}{c|}{\(K=3\)} & \multicolumn{4}{c}{\(K=4\)}\\
			 & Cluster 1 & Cluster 2 & Cluster 1 & Cluster 2 & Cluster 3 & Cluster 1 & Cluster 2 & Cluster 3 & Cluster 4\\
			\hline
			\multirow{2}{*}{STO1} & 0 s –-  & 2 s –- & 0 s –- & 2 s -- & 435 s –- & 0 s –- & 15 s –- & 434 s –- & 729 s –- \\
			 & 3479 s & 3333 s & 3393 s & 1747 s & 3479 s  & 1143 s & 712 s & 3393 s & 3479 s \\
			\multirow{2}{*}{STO2} & 0 s –- & 1387 s –- & 0 s -– & 214 s -– & 1387 s -– & 0 s -– & 214 s -– & 217 s -– & 1929 s -– \\
			 & 2044 s & 3524 s & 213 s & 2094 s & 3524 s & 213 s & 1072 s & 2601 s & 3524 s \\
			\multirow{2}{*}{STO3} & 0 s -– & 914 s –- & 0 s –- & 683 s –- & 1666 s –- & 0 s -– & 337 s -– & 792 s -– & 1666 s –- \\
			 & 989 s & 3581 s & 736 s & 1742 s & 3581 s & 374 s & 809 s & 1744 s & 3581 s \\
			\multirow{2}{*}{STO4} & 0 s -– & 897 s –- & 0 s –- & 666 s –- & 1338 s -– & 0 s –- & 220 s -– & 788 s -- & 1407 s -– \\
			 & 1043 s & 3512 s & 580 s & 1377 s & 3512 s & 227 s & 796 s & 1415 s & 3512 s \\
		\end{tabular}
	\end{ruledtabular}
\end{table*}

\begin{figure*}
	\includegraphics{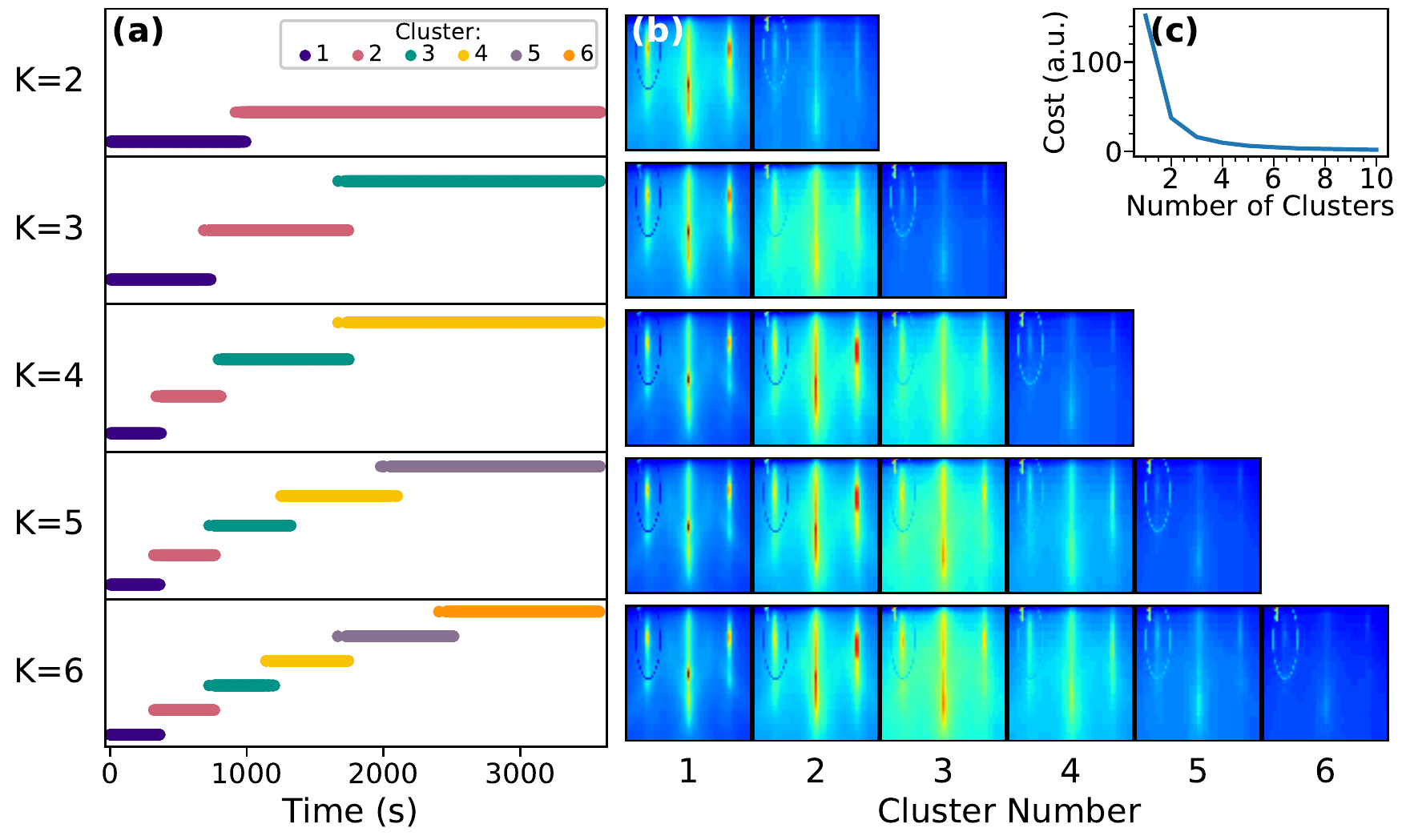}
	\caption{\label{STO3}(a) K-means clustering for \(K=1-8\) clusters, (b) the mean image of each cluster, and (c) the k-means minimization function plotted for each value of \(K\) for STO3.}
\end{figure*}

For the non-stoichiometric samples, STO3 and STO4, the groupings in the evolution of the growth are more distinct. Clustering the frames into 3 or more groups for each non-stoichiometric sample leaves a final phase that suggests a stable growth with little evolution in the RHEED pattern after 1666 seconds for STO3 and $\sim$1400 seconds for STO4 (Table~\ref{STOClusterDuration}). A more in-depth look at clustering STO3 into up to 8 clusters (Fig.~\ref{STO3}(a)) demonstrates that most cluster divisions occur early on in the growth, suggesting that changes in the surface evolution reach a zenith in the initial stages of growth until a steady-state phase is achieved in the final portion of the growths. The mean images for each cluster (Fig.~\ref{STO3}(b)) demonstrate the same basic RHEED pattern evolution in each clustering—as material is deposited on the surface, there is an increase in RHEED pattern intensity with a gradual smearing of the spots from the substrate into streaks, and then a fade away of the streak intensity. The first cluster in all cases features a weak 2$\times$ reconstruction that fades away into a 1$\times$ by the second or third cluster, which is consistent with a Ti-rich surface \cite{kajdos2014surface, jalan2009growth, fisher2008stoichiometric}.

Even for samples STO3 and STO4, which form more clearly defined clusters, the number of clusters that is most appropriate is not clearly defined. As naive k-means clustering relies on a user to define the number of clusters that should be grouped, the algorithm provides no obvious ``natural'' grouping. A minimization parameter, or cost function, for the algorithm is plotted in Fig.~\ref{STO3}(c) as a function of the number of clusters. In the algorithm, the cost function \(J\) for a given number of clusters \(K\) is defined as the sum of the distances between each frame that is grouped into the cluster and the cluster centroid (or ``mean'' image for each cluster):
\begin{equation}
J_k=\frac{1}{m}\sum_{i=1}^{m}\|x^{(i)} - \mu_{c^{i}}\|
\end{equation}
where \(m\) is the number of frames in the video, \(c^{i}\) is the index of the cluster to which image \(x\) is assigned, and \(\mu_{c^{(i)}}\) is the cluster centroid of the cluster to which \(x\) has been assigned. Thus, the ``optimal'' number of clusters will always occur when the number of clusters is equal to the total number of frames, i.e. the centroid of each cluster will match with a distinct frame and which will provide a total cost of zero. The cost function will also monotonically decrease as the number of clusters is increased, although looking for an ``elbow'' in the curve can hint to an optimal number of clusters to use for analysis. In most of the samples studied here, the cost function is combined with previously known factors about the growth and the mean images for each cluster to determine the most useful number of groupings.

\subsection{STO on Scandate Substrates}
\begin{figure*}
	\includegraphics{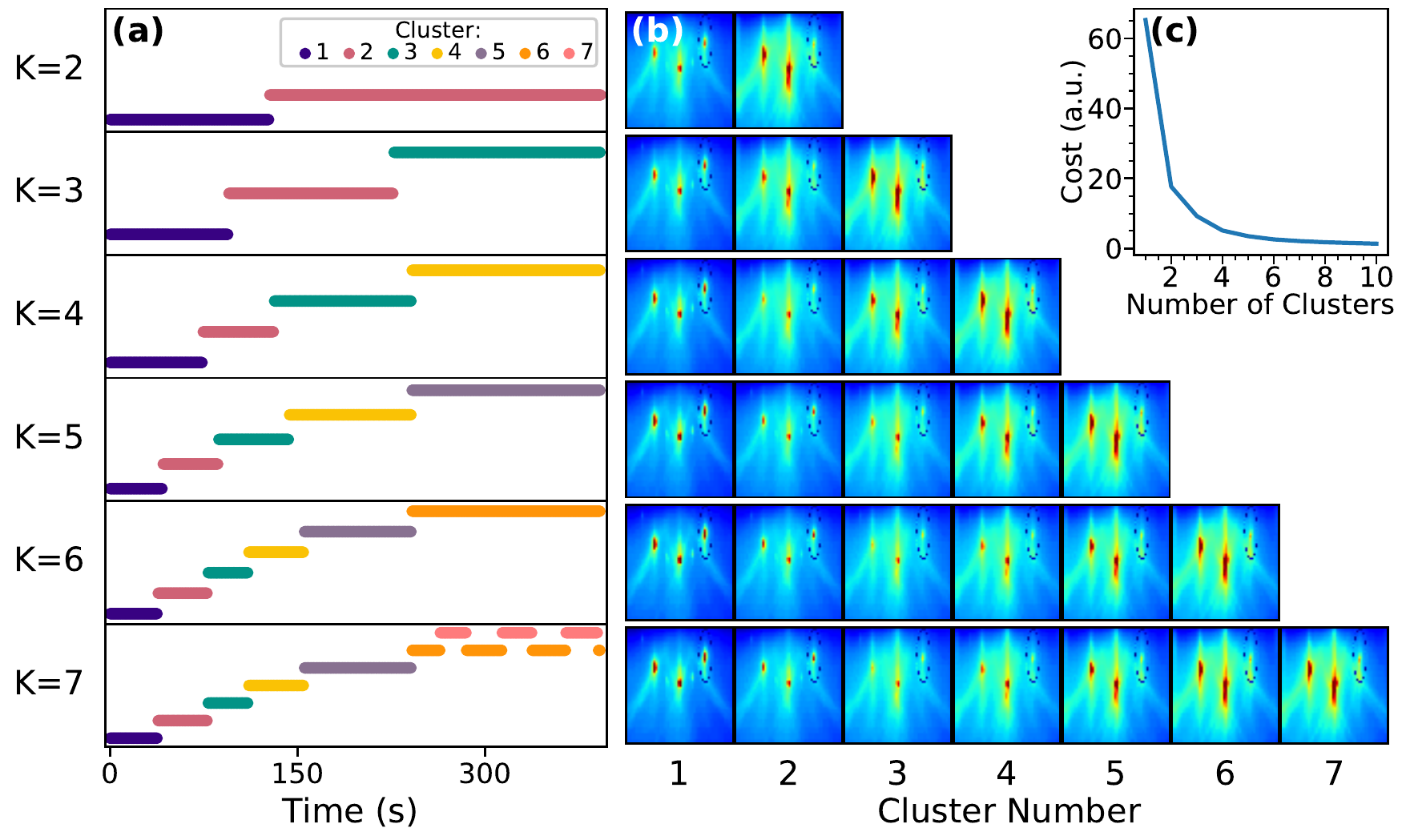}
	\caption{\label{STOTSOClustering}(a) K-means clustering up to \(K=7\) for STO on TSO, along with (b) the mean representative images in each cluster and (c) the k-means minimization function plotted for each value of \(K\).}
\end{figure*}

\begin{figure*}
	\includegraphics{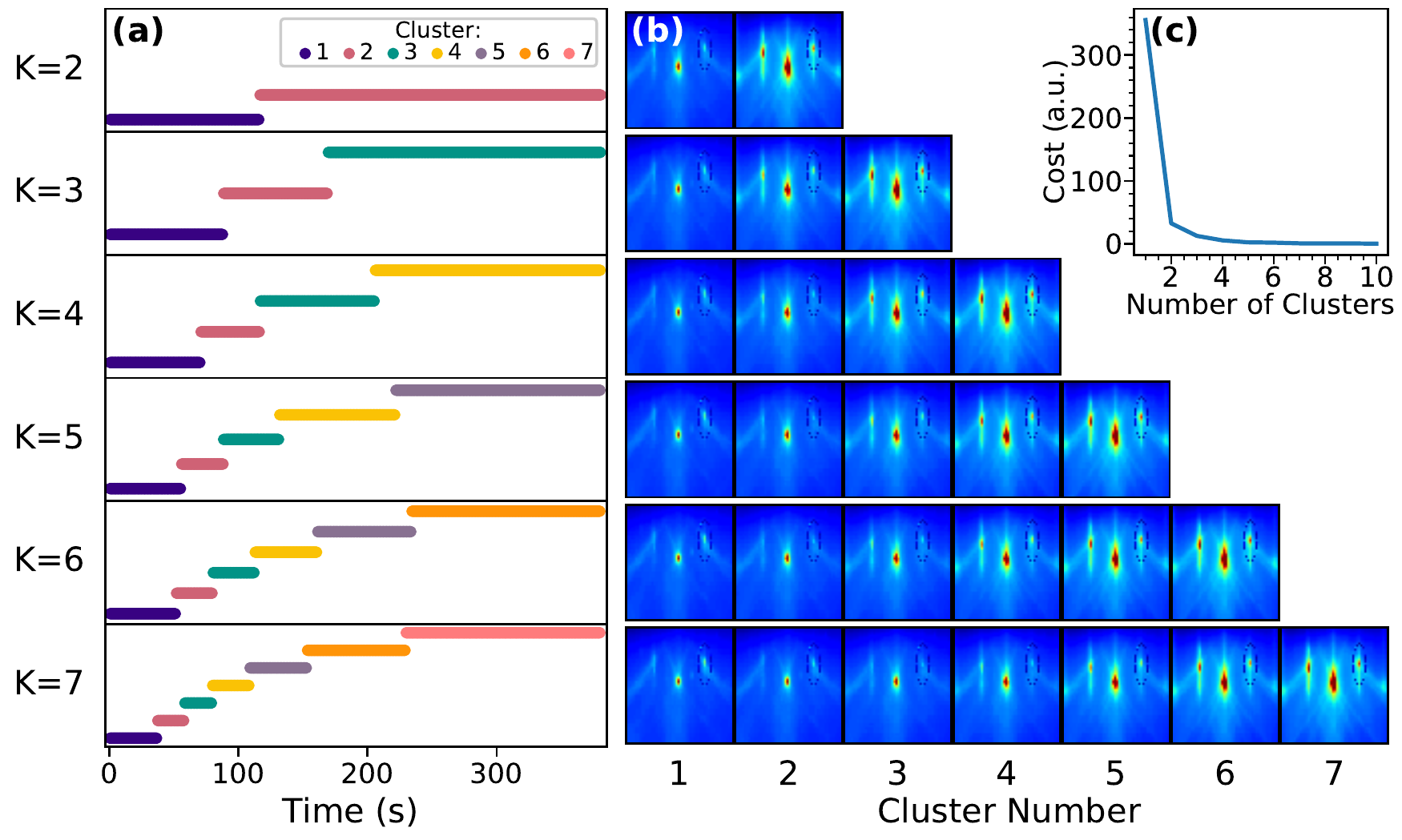}
	\caption{\label{STOGSOClustering}(a) K-means clustering up to \(K=7\) for STO on GSO, along with (b) the mean representative images in each cluster and (c) the k-means minimization function plotted for each value of \(K\).}
\end{figure*}

\begin{figure}
	\includegraphics{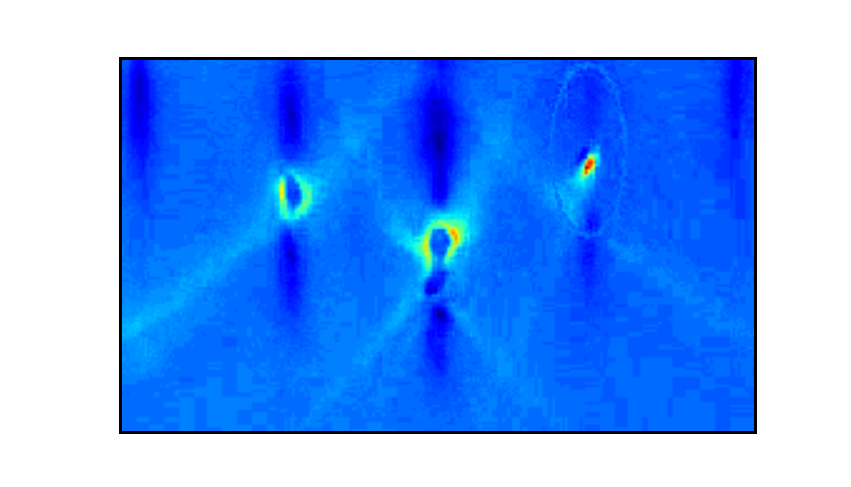}
	\caption{\label{ClusterDiff}The difference between the mean images for clusters 6 and 7 for the STO film grown on a TSO substrate.}
\end{figure}

\begin{figure*}
	\includegraphics{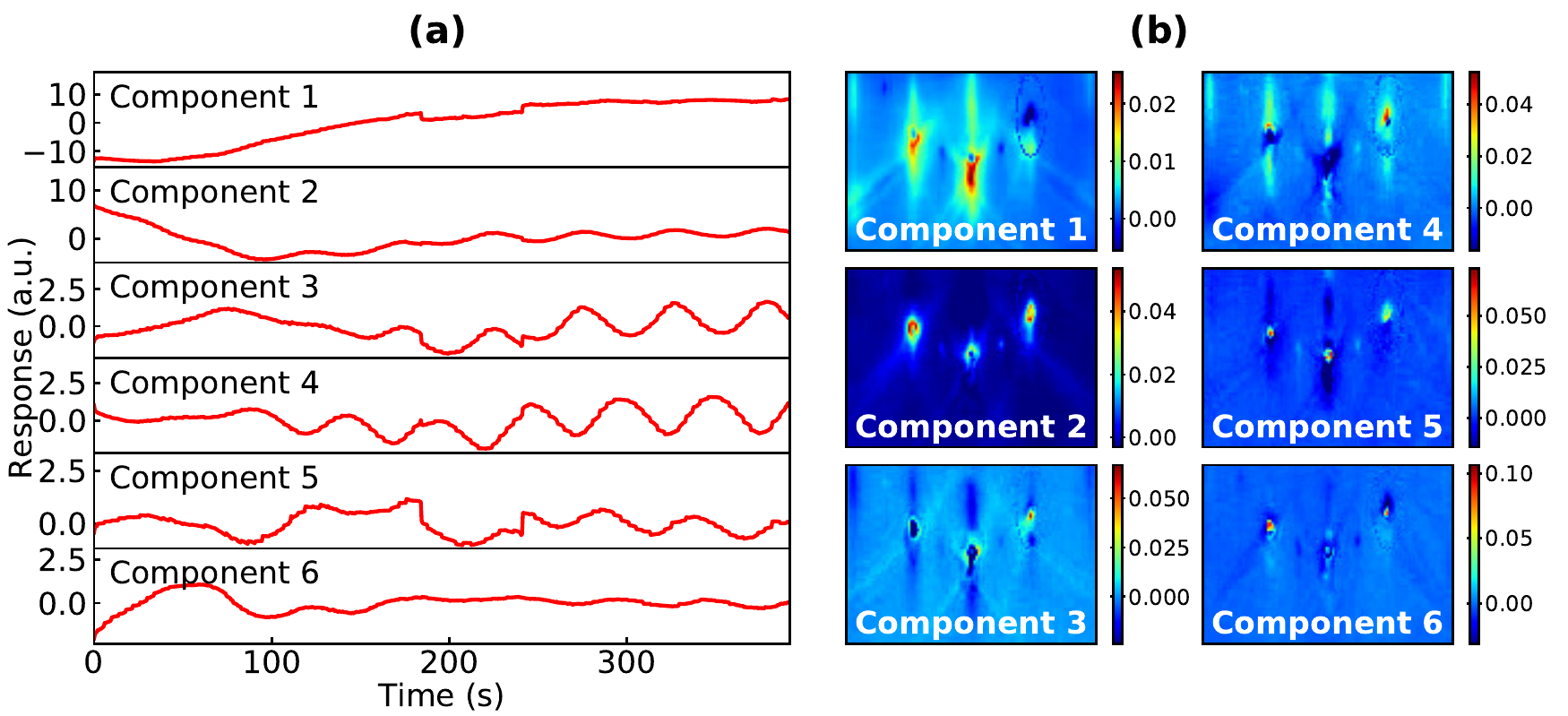}
	\caption{\label{STOTSO_PCA}(a) The loadings plotted as a function of time and (b) the corresponding first six principle components resulting from PCA for the STO film grown on a TSO substrate.}
\end{figure*}

STO films grown using similar stoichiometric conditions to those described in the previous section were grown on $\langle$110$\rangle$ GdScO$_{3}$ (GSO) and TbScO$_{3}$ (TSO) substrates, which provide a pseudocubic (pc) (001) surface mesh for growth. The results of k-means clustering the RHEED video taken along the [110]$_{pc}$ azimuths from each sample are presented in Figs.~\ref{STOTSOClustering} and \ref{STOGSOClustering}, clustered up to \(K=7\). The growth on the GSO lasted for $\sim$360 seconds, and the growth on TSO lasted for $\sim$400 seconds, so the analysis includes some frames of the static film after the deposition had concluded. In each case, the mean images of each sample (Figs.~\ref{STOTSOClustering}(b) and \ref{STOGSOClustering}(b)) and the plot of the cost function (Figs.~\ref{STOTSOClustering}(c) and \ref{STOGSOClustering}(c)) indicate little utility in clustering beyond \(K=2\). The RHEED from both films seem to indicate the steady formation of stoichiometric STO, including the development of strong Kikuchi bands and an increase in the RHEED intensity. There are very few obvious visual differences between the mean RHEED pattern in each cluster when using higher values of \(K\) beyond an increase in intensity of the streaks in each image and the loss of the half order spots on the scandate substrates, particularly in the case of the STO film grown on GSO. The cost function for this film has an obvious elbow at \(K=2\), and increasing the number of clusters doesn’t seem to have any strong physical meaning aside from indicating an increase in RHEED pattern intensity. Indeed, values of \(K\) greater than or equal to 5 for the STO film on TSO include the frames after the growth cutoff as its own grouping, while clustering using \(K=7\) begins to form distinct clusters from the intensity changes due to RHEED oscillations. The difference between the mean images for clusters 6 and 7 is displayed in Fig.~\ref{ClusterDiff}, which shows that the predominant difference between the images in each cluster is the intensity of the ``halo'' surrounding each RHEED spot and the Kikuchi bands. The oscillation of the Kikuchi features is consistent with previous reports that the periodic RHEED intensity oscillations are strongly affected by the presence of Kikuchi bands \cite{zhang1987effects, shin2007phase}. The timing of the oscillations between clusters averages 49 seconds per cycle, or 0.02 Hz, which is the same as the periodicity extracted from the loading plot in Fig.~\ref{STOTSO_PCA}. 

\begin{figure*}
	\includegraphics{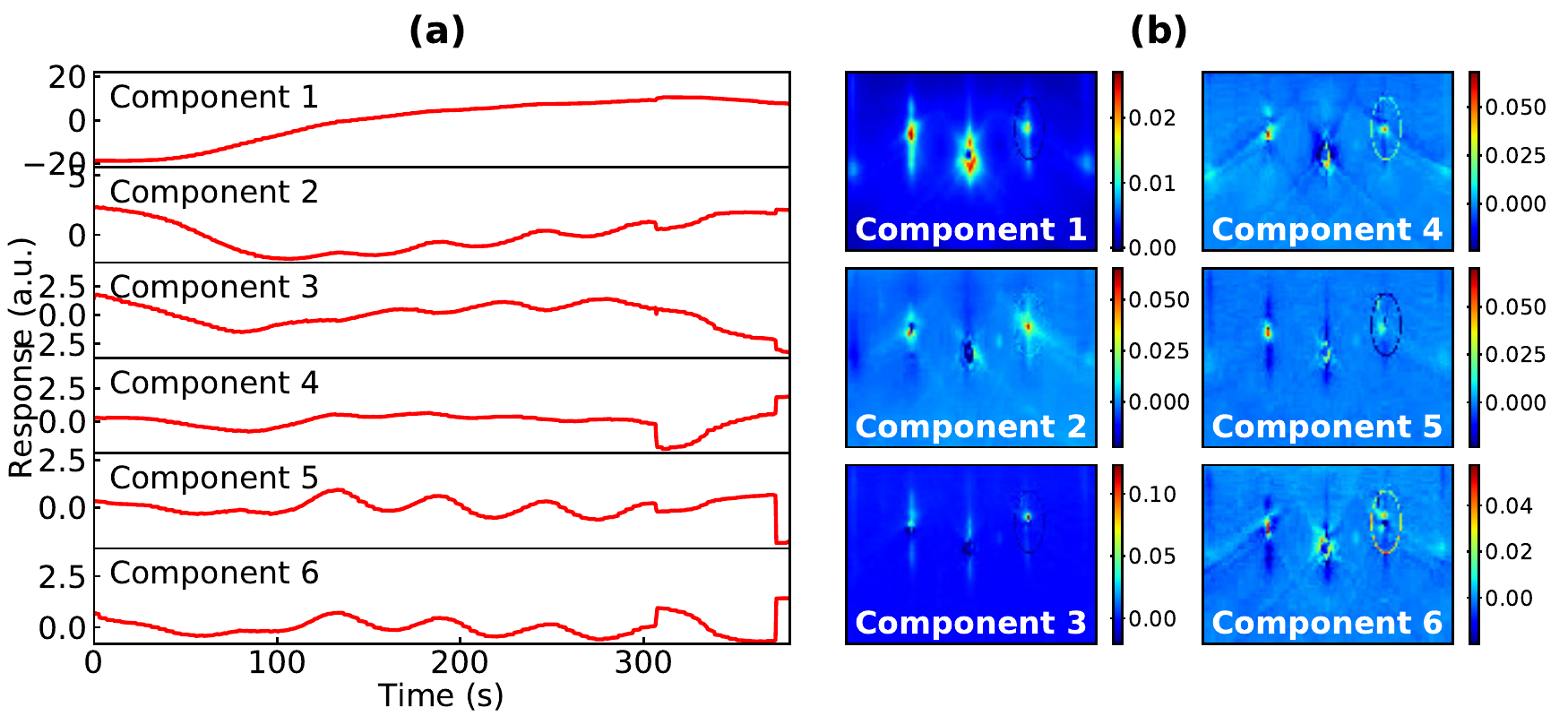}
	\caption{\label{STOGSO_PCA}(a) The loadings plotted as a function of time and (b) the corresponding first six principle components resulting from PCA for the STO film grown on a GSO substrate.}
\end{figure*}

One possible interpretation of the clear boundary between the two clusters for \(K=2\) (at 128 seconds for the STO film grown on TSO and at 117 seconds for the STO film grown on GSO) would be a transition in growth mode. The mean images for the cluster centroids for the film grown on TSO (Fig.~\ref{STOTSOClustering}(b)) reveal that the initial RHEED features smear out and form a modulated (00) streak, indicating a transition from a relatively smooth surface to a multilevel stepped surface, although this change appears gradual rather than occurring at a single boundary. Plotting the loadings as a function of time for each film (Figs.~\ref{STOTSO_PCA}(a) and \ref{STOGSO_PCA}(a))) captures RHEED oscillations for each sample occurring at $\sim$0.02 Hz, indicating that the film entered a layer-by-layer growth mode at approximately the same times as the cluster boundary formed for each sample when clustered with \(K=2\). There is no significant shift in the spacing between RHEED streaks over time that would indicate relaxation has occurred, although with lattice mismatch of $\sim$40\% for both STO films it is expected that each film would relax relatively quickly. A ``streaky'' pattern is generally observed within the first few unit cells of the growth process, as can be seen in the time-dependent loadings for principle components 3-5 and the streaks in the corresponding components in Fig.~\ref{STOTSO_PCA}. These streaky features were described as corresponding to imperfect layer-by-layer growth by Vasudevan et al \cite{vasudevan2014big} and the positive amplitude decreases after the first few unit cells, suggesting that the growth mode transitions in to a cleaner layer-by-layer mode within about 4 unit cells.

\subsection{BSO on STO Substrates}
The k-means clustering analysis was performed on BSO films grown at the University of Minnesota on $\langle$100$\rangle$ STO substrates using a hybrid MBE reactor equipped with hexamethylditin, (CH$_{3}$)$_{6}$Sn$_{2}$ (HMDT) as a tin precursor and a Ba effusion cell \cite{prakash2015hybrid}. The substrate (thermocouple) temperature was held fixed at 900$^{\circ}$C and the BSO film was grown for 60 minutes ($\sim$60 nm), although video was only processed for the first 10 minutes. The ratio of HMDT to Ba beam equivalent pressure (BEP) as measured by a beam flux monitor (BFM) is 18.1, and the film appears to be stoichiometric.

\begin{figure*}
	\includegraphics{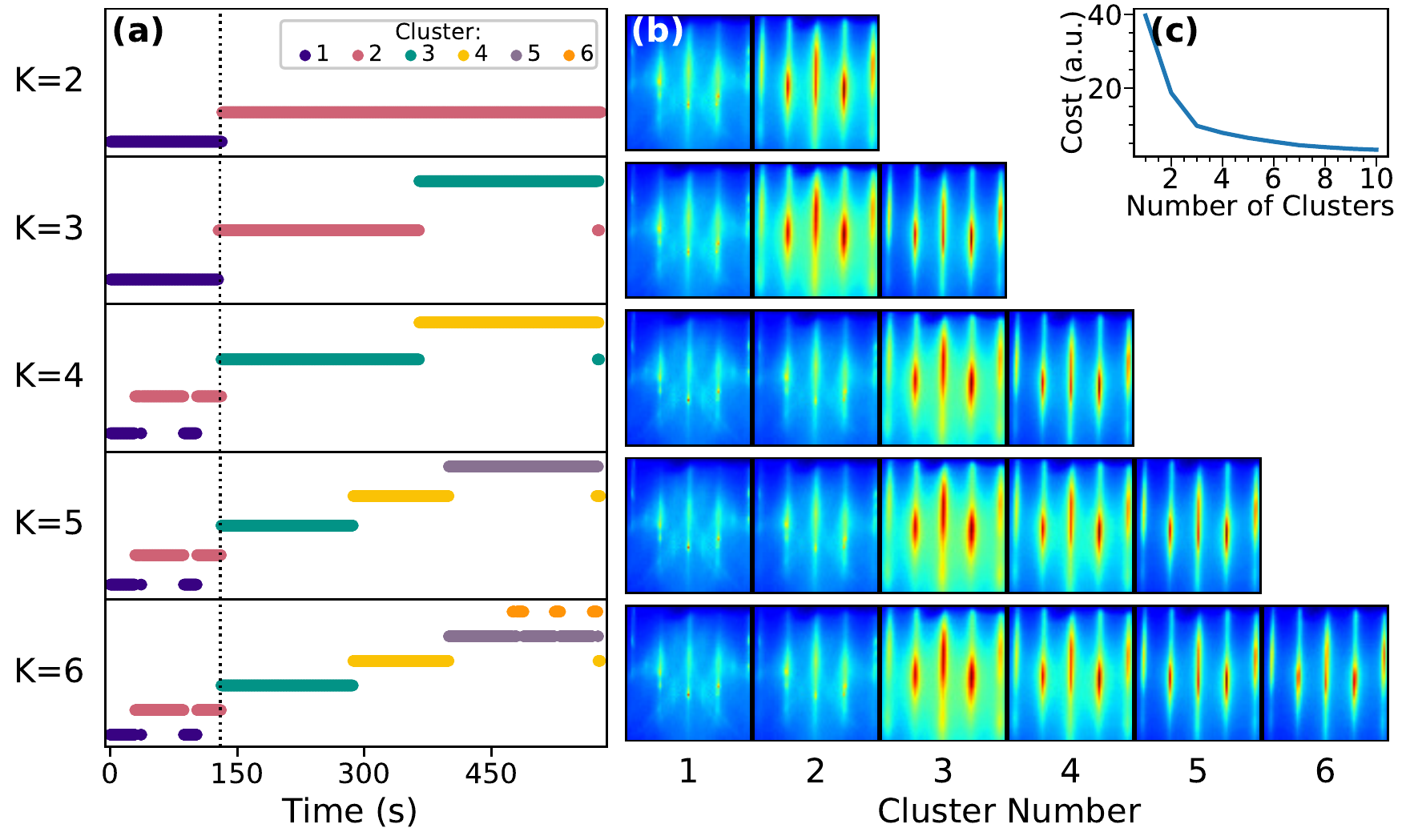}
	\caption{\label{BSOClustering}(a) K-means clustering up to \(K=6\) for BSO on STO, along with (b) the mean representative images in each cluster and (c) the k-means minimization function plotted for each value of \(K\). The dotted line in (a) emphasizes the consistency of the cluster boundary at 130 seconds in all groupings.}
\end{figure*}

\begin{figure*}
	\includegraphics{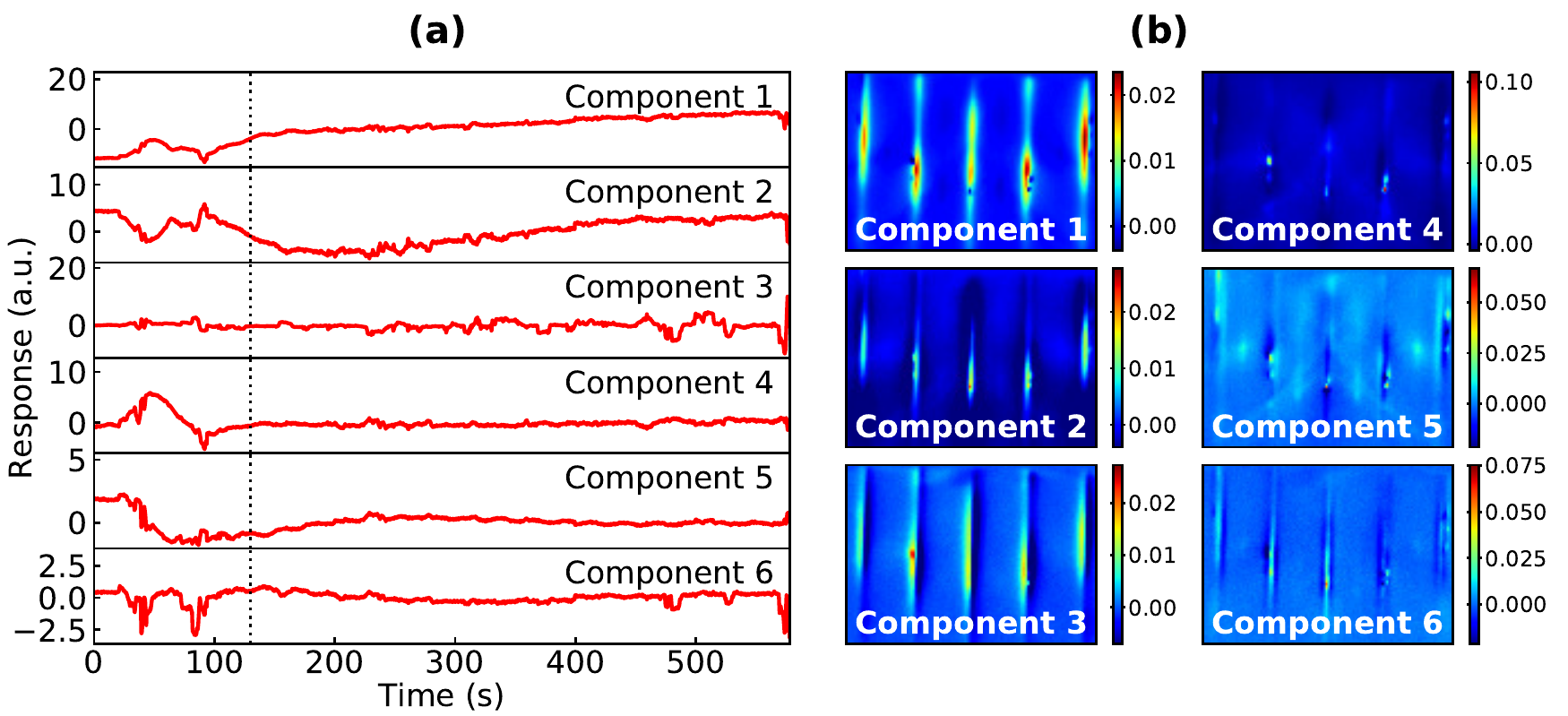}
	\caption{\label{BSO_PCA}(a) The loadings plotted as a function of time and (b) the corresponding first six principle components resulting from PCA for BSO grown on an STO substrate. The dotted line in (a) at 130 seconds indicates the primary cluster boundary for the sample (Fig.~\ref{BSOClustering}(a)).}
\end{figure*}

The progression of clustering charts the progression of the RHEED images along the [100] azimuth from the spots of the STO substrate in the early stages of growth to a streakier film pattern. The addition of clusters predominantly seems to mark changes to the intensity of the RHEED pattern rather than discernable changes in the pattern itself. Clustering the frames into more than three groups (Fig.~\ref{BSOClustering}(a)) begins to elucidate intensity fluctuations visible in the loadings (Fig.~\ref{BSO_PCA}). With \(K=4\) and greater, gaps occur in the first two clusters corresponding to changes in the loading response within the first 100 seconds. In the \(K=6\) grouping, cluster 6 directly corresponds to the response spikes visible in principle components 3 and 6. Regardless of \(K\) value, however, there is a boundary between cluster groupings after 130 seconds, suggesting a transition in the film growth at this time.

As a concrete measurement of the differences in the mean images before and after this boundary at 130 seconds, the in-plane lattice parameter was calculated for each of the mean images in the cluster using the separation of the diffraction streaks \cite{whaley1990relaxation}. Using the initial substrate pre-growth peak spacing as a reference, we calculate a lattice parameter of 3.9 {\AA} for all mean cluster images before 130 seconds, and a lattice parameter of 4.1 {\AA} for all mean cluster images occurring after 130 seconds for the \(K=2\) and \(K=3\) groupings, which is consistent with the lattice parameter of a = 3.905 {\AA} for STO and a reported in-plane lattice parameter of a = 4.107 {\AA} for BSO grown on an STO substrate \cite{kim2012high, kim2018laino3}. The abrupt change in the in-plane lattice parameters before and after 130 seconds indicates that film relaxation occurred at this boundary. 

Although RHEED oscillations are hard to discern in the loadings plotted as a function of time, most variance in the loading response occurs in the early stages of the growth. Given a measured growth rate of 54 nm/hour (from dividing the overall growth time by the measured thickness from XRD fringes), the periodicity of RHEED oscillations would be expected to be 27 seconds for one unit cell (u.c.) of film coverage. Small oscillations with a periodicity of 27 seconds are visible in the loadings for most components in Fig.~\ref{BSO_PCA}(a), although they are often dwarfed by noise attributed to vibrations in the system. Furthermore, previous analysis by Prakash et al \cite{prakash2015hybrid} has found an approximate strain relaxation thickness of $\sim$1 nm for similar BSO films on STO. The hump in the response of principle component 4 within the first 130 seconds of growth mirrors the RHEED intensity pattern characteristic of strain relaxation, suggesting strain relaxation occurring in the film at $\sim$2 nm. This boundary coincides with the transition in clustering that occurs at the same time in Fig.~\ref{BSOClustering}, indicating that the boundary forms as a result of strain relaxation in the film. This result is consistent with the transition from spottier patterns in the mean images from k-means (cluster 1 for \(K=2, 3\) and clusters 1 and 2 for \(K>4\), Fig.~\ref{BSOClustering}(b)) to a streakier pattern in the latter part of the growth. The gradual compression of the streaky features in the mean images in Fig.~\ref{BSOClustering}(b) and the oscillations in the loadings with periodicity of 27 seconds in Fig.~\ref{BSO_PCA}(a) may indicate that the film surface is smoothing after a few layers of pseudomorphic growth and returning to a layer-by-layer growth mode. 

\subsection{LNO on LAO Substrates}
\begin{figure*}
	\includegraphics{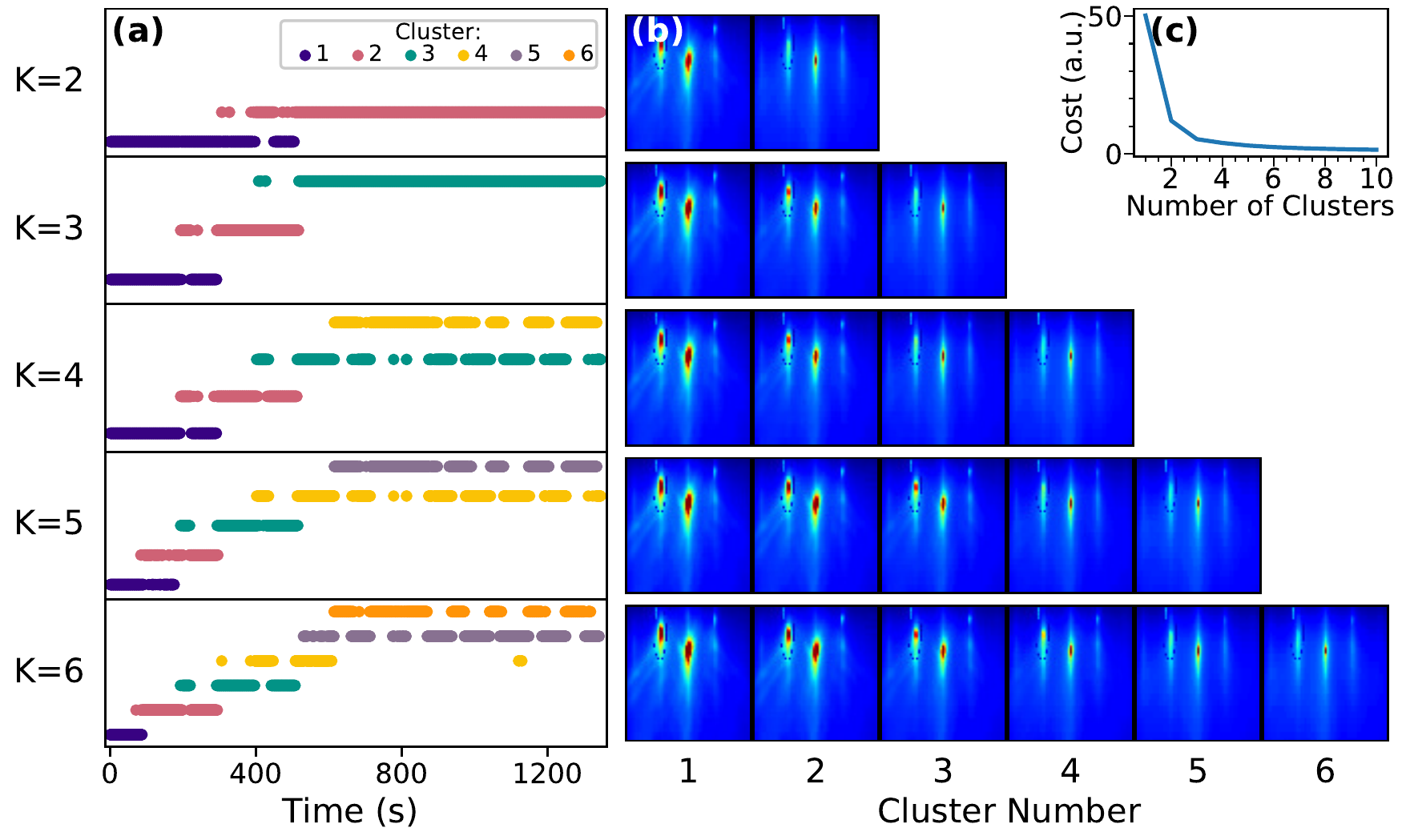}
	\caption{\label{LNOClustering}(a) K-means clustering up to \(K=6\) for LNO on STO, along with (b) the mean representative images in each cluster and (c) the k-means minimization function plotted for each value of \(K\).}
\end{figure*}

To examine the sensitivity of the algorithm to shuttered growth, an LNO film was grown on a (100)$_{pc}$ LAO substrate using shuttered deposition, in which the LaO flux was alternated with the NiO$_{2}$. The shutter sequence (Fig.~\ref{LNOClustering}(a)) was alternated between the Ni and La sources every 30 seconds, with a 45 second anneal period between deposition layers and a one second transition between each shutter change. Both the La and Ni were deposited from standard effusion cells, and the oxygen was supplied with an RF plasma source. The substrate was held at a constant temperature of 600$^{\circ}$C with a constant background oxygen flow rate of 2.5 sccm, producing a chamber pressure of $\sim$2 $\times$ 10$^{-5}$ Torr. LNO is a challenging material to synthesize by MBE due to the propensity to form oxygen vacancies in the oxygen pressure regimes typically accessible in an MBE chamber ($P_{O_{2}} < 10^{-5}$ Torr). Thus, a shuttered growth scheme and annealing step have been employed by some groups to more fully oxidize the material \cite{king2014atomic}.

K-means clustering is not particularly revealing with this film, as the clusters primarily track the intensity oscillations of the RHEED pattern as the shutter pattern shifts (Fig.~\ref{LNOClustering}), although the periodicity of the oscillations in the k-means clustering does not precisely align with the shuttering sequence. In all cluster groupings, a loose boundary appears at $\sim$500 seconds into the growth, or after the deposition of 4 u.c. of LNO. The mean images for each cluster indicate a dissolution of the Kikuchi bands that appear in the earlier stages of the growth (appearing in cluster 1 for \(K=2\), clusters 1 and 2 for \(K=3, 4\), and clusters 1-3 for \(K=5, 6\)) as the RHEED transitions into a streakier pattern. Note that the Kikuchi bands are not symmetric in the early mean images, indicating that the substrate rotation is slightly off the [110] orientation. The dispersed nature of the clusters for all values of \(K\), however, indicates that this is more of a gradual process than an abrupt change at a specific timestep.

\begin{figure*}
	\includegraphics{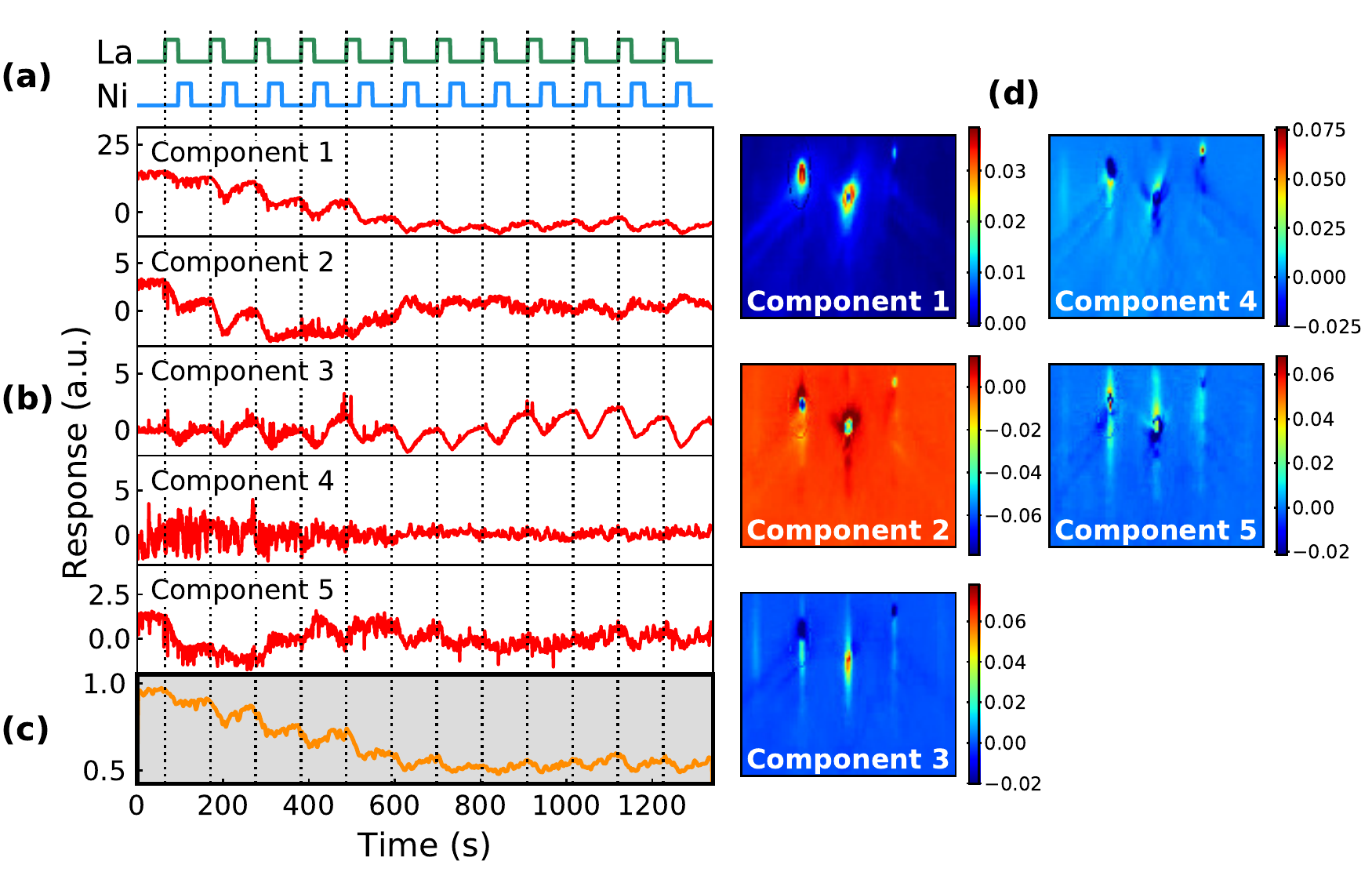}
	\caption{\label{LNO_PCA}(a) The shutter sequence of the LNO growth on LAO. The La shutter (green) is opened for 30 seconds against a constant oxygen background flux to deposit LaO, followed opening the Ni shutter (blue) for 30 seconds. The film is then annealed for 45 seconds after the NiO2 deposition before restarting the sequence. The cycle is repeated 12 times during the growth, with the boundaries between the shutter cycle demarcated with the vertical dotted lines. (b) The loadings plotted as a function of time and (c) the mean RHEED intensity of the [-10] streak are plotted against the shutter sequence cycle, so as to highlight the periodic changes in the loading/RHEED pattern corresponding to the shutter sequence of the growth. (d) The first five principle components resulting from PCA. }
\end{figure*}

The response of the loadings over time (Fig.~\ref{LNO_PCA}(b)) precisely aligns with the 107 second periodicity of the mean intensity of the [-10] RHEED streak (Fig.~\ref{LNO_PCA}(c)). The loading response over time for principle components 1 and 3 mirrors the mean RHEED intensity fairly well, as the intensity increases during the NiO$_{2}$ deposition, wanes during the LaO deposition, and flattens during the anneal phase. This alignment between the periodicity of the RHEED intensity oscillations and loadings demonstrates that the loadings derived from PCA contains at least as much information as traditional RHEED oscillations derived from tracking the mean intensity of a user-specified spot during film growth.

The principle components give a few clues as to the nature of the transition present in the k-means clustering after 4 u.c. Principle component 1 broadly represents the Kikuchi bands as well as the halo 
around the specular spots in the RHEED pattern. There is a notable gap in the center of the RHEED spots in component 1, indicating that there is almost no contribution in the center of the spot from this feature. Component 1 dominates in the early part of the growth, but trends towards and then begins to oscillate around zero after the deposition of 5 u.c. In contrast, the oscillations displayed by component 3 (which represents the streakier features prominent in the latter part of the growth) are steady throughout the growth, but there a significant decrease in noise with a corresponding increase in amplitude of the oscillations after 4 u.c. have been deposited. There is a similar decrease in noise after 4 u.c. in component 4, which we interpret to largely represent noise present in the chamber during growth. There is a small vibration present in the substrate holder during RHEED growth, and the Kikuchi bands can be seen increasing and decreasing in intensity throughout the growth. The intensity of the Kikuchi bands is dependent on crystal orientation, so small fluctuations would cause interference in the bands. The decrease in the amplitude of the noise at 4 u.c. implies that contribution of the Kikuchi bands to the overall RHEED pattern at this point in the growth has mostly decreased, as we assume the vibrational noise is present throughout the growth.

The presence of RHEED oscillations throughout the duration of the growth indicates that the LNO film is being deposited in a layer by layer growth mode. LNO and LAO are almost lattice-matched with bulk pseudocubic lattice constants of 3.84 {\AA} and 3.83 {\AA} \cite{wrobel2017comparative}, respectively, so it is unsurprising that there is no measurable change in the in-place lattice parameter of the course of the growth and that the LNO film remain metamorphic. The quality of the crystalline surface is difficult to determine from PCA or k-means clustering, however. The transition point at 4 u.c. that k-means clustering largely seems to indicate is due to a decrease in both the intensity of the Kikuchi bands and from the overall specular pattern. The transition in the growth apparent in all k-means groupings at 4 u.c. may be indicative of an overall decrease in quality of the crystalline surface, or transition from an atomically flat surface to a more terraced, rougher surface. 

From the observed oscillations in components 1 and 3 during the shuttering and annealing sequence, we observe that the growth of the LaO layer reduces the intensity of the streak pattern (imperfect layer-by-layer growth, component 3). Conversely, growth of the NiO$_{2}$ layer strengthens both the ideal layer-by-layer growth (component 1) and negates the changes to component 3 that occur due to the growth of the LaO layer. Finally, we also observe that the annealing step for the final 45 seconds of a cycle continues to strengthen the ideal layer-by-layer spot pattern in component 1 even when all shutters are closed. This suggests that the annealing step is important for crystallization of a smooth film surface and more complete oxidation of the LNO film, as others have observed empirically \cite{king2014atomic}.

\section{Conclusions}
In summary, methods to analyze an entire RHEED data set were applied to different types of perovskite film growths using PCA and k-means clustering. We have specifically applied these approaches to understand the stoichiometry of homoepitaxial SrTiO$_{3}$ thin films, surface evolution in heteroepitaxial SrTiO$_{3}$ films grown on GdScO$_{3}$ and TbScO$_{3}$, strain relaxation in BaSnO$_{3}$ films grown on SrTiO$_{3}$, and surface crystallinity during a shuttered growth for LaNiO$_{3}$ films. Compression of the data using PCA and the analysis of the loadings and principle components produced may provide an alternative to in situ monitoring of RHEED oscillation intensity as the same intensity oscillations appear in the loadings produced over time. Information contained within the principle components can provide additional insights into the physical significance of RHEED oscillations and can be used to understand surface evolution during the growth process. K-means clustering may provide information about transitions in the growth modes at precise times in the growth, although care must be taken to consider the appropriate number of clusters to use during analysis. Given the ubiquity of RHEED data acquisition during film growth and the rising use of big data analytics, we suggest that video archival of the entire RHEED image sequence instead of just intensities of regions of interest can provide significant additional information about the materials being synthesized. For this reason, we have made the source code used in this work freely accessible for others to analyze a wide range of materials systems.

\section{Supplemental Material}
See supplemental material for more detail on the principle component analysis and k-means clustering as applied to RHEED video, details on each film growth, and a link to the source code for use by other researchers.

\vfill
\section{Acknowledgments}
The authors would like to acknowledge the Auburn University Hopper Cluster for support of this work. Scandate substrates were provided through the National Science Foundation [Platform for the Accelerated Realization, Analysis, and Discovery of Interface Materials (PARADIM)] Materials Innovation Platform under Cooperative Agreement No. DMR-1539918. S. Thapa and S. R. Provence gratefully acknowledge support from the Auburn University Department of Physics. R. J. Paudel and R. B. Comes gratefully acknowledge support for the LaNiO$_{3}$ work from NSF-DMR-1809847. Work at the UMN involving thin film growth and characterization was supported by the U.S. Department of Energy through DE-SC0020211. 

\bibliographystyle{apsrev4-2}
\bibliography{Bibliography}
\end{document}